\documentclass{rmf-d}
\usepackage{nopageno,rmfbib,multicol,times,epsf,amsmath,amssymb,cite}
\usepackage[T1]{fontenc} 
\usepackage[]{caption2}
\usepackage{graphicx}
\usepackage{blindtext}
\usepackage{color}
\usepackage{hyperref}
\def\rmfcornisa{
}

\def\rmfcintilla{{\it Rev.\ Mex.\ Fis.\/} {\bf ??} (*?*) (????) ???--???}
\begin{document}

%
%
\title{Core-corona approach to describe hyperon global polarization in semi-central relativistic heavy-ion collisions
\vspace{-6pt}}

\author{Alejandro Ayala$^{(1,2,3)}$, Isabel Domínguez$^{(4)}$, Ivonne Maldonado$^{(5)}$, Mar\'{\i}a Elena Tejeda-Yeomans$^{(6)}$\\
(for the MPD Collaboration)}

\address{$^{(1)}$Instituto de Ciencias Nucleares, Universidad Nacional Aut\'onoma de M\'exico, Apartado Postal 70-543, CdMx 04510, México.\\
$^{(2)}$Centre for Theoretical and Mathematical Physics, and Department of Physics, University of Cape Town, Rondebosch 7700, South Africa.\\
$^{(3)}$Departamento de F\'isica, Universidade Federal de Santa Maria, Santa Maria, RS 97105-900, Brazil.\\
$^{(4)}$Facultad de Ciencias F\'isico-Matem\'aticas, Universidad Aut\'onoma de Sinaloa, Avenida de las Am\'ericas y Boulevard Universitarios, Ciudad Universitaria, C.P. 80000, Culiac\'an, Sinaloa, México.\\ $^{(5)}$Joint Institute for Nuclear Research, Dubna, 141980, Russia.\\ $^{(6)}$Facultad de Ciencias - CUICBAS, Universidad de Colima, Bernal D\'iaz del Castillo No. 340, Col. Villas San Sebasti\'an, 28045 Colima, México.}

%
%

\author{ }
\address{ }
\author{ }
\address{ }
\maketitle
%
%
\begin{abstract}
\vspace{1em} 
%
%
We report on the core-corona model developed to describe the main features of hyperon global polarization in semicentral relativistic heavy-ion collisions as a function of the collision energy. We first neglect the contribution to polarization from hyperons produced in the corona. In this scenario, the global polarization turns out to be described by a delicate balance between the vorticity-to-spin transferring reactions in the core and the predominance of corona over core matter at low energies. We show how this last feature provides a key ingredient missing in our original model that helps to better describe the excitation function for $\Lambda$ and $\overline{\Lambda}$ global polarization. To improve the description, we then introduce the contribution to the global polarization coming from the transverse polarization of $\Lambda$s produced in the corona, which is hereby assumed to be similar to the well-known polarization produced in p + p reactions. The results show a small positive contribution to the global polarization, however they are not yet conclusive due to the small size of the MC sample used in the analysis.
\end{abstract}
\begin{multicols}{2}

\section{Introduction}

Studies of hyperon polarization have been part of the standard tool kit of high-energy proton and heavy-ion collisions aiming to better characterize the role played by spin in strong interactions. In fact, the difference of hyperon polarization in central vs. peripheral heavy-ion collisions was put forward some time ago as a means to identify the production of QGP formation~\cite{Ayala:2001jp}. Nowadays, the study of hyperon global polarization is an integral part of the QGP characterization program, linked to QGP properties such as viscosity, vorticity and flow. Recent results on global hyperon polarization, reported by STAR and HADES, for mid-central heavy-ion collisions, show that the global polarization increases with the decrease of collision energy, and that this effect is larger for $\bar{\Lambda}$s than for $\Lambda$s~\cite{STAR:2017ckg,STAR:2021beb,HADES:2022enx}.

In recent works~\cite{Ayala:2020soy,Ayala:2021xrn} we have developed a two-component model to describe the $\Lambda$ and $\bar{\Lambda}$ polarization. The properties of the polarization excitation functions are linked to the relative abundance of $\Lambda$s coming from the two distinct regions of the system created in the collision: a low-density corona and a high-density core. Core-corona approaches have been previously used in different contexts, for instance, to explain the production of strange particles, the dependence of elliptic flow and the average of the mean transverse momentum as functions of centrality at RHIC and SPS energies~\cite{Becattini:2008ya,Aichelin:2010ed}.

The model does a good average description of the global polarization obtained in semi-central collisions of heavy systems, as a function of the collision energy. However, as discussed in Ref.~\cite{Ayala:2022yyx}, for the description of the polarization arising in collisions of smaller systems, such as Ag + Ag at $\sqrt{s_{NN}} = 2.55$ GeV, in the 10-40\% centrality class, or for large systems such as Au + Au at $\sqrt{s_{NN}}= 3$ GeV, for centralities larger than 40\%, the model has limitations. To improve, we have argued the necessity to include in the description the polarization of $\Lambda$s and $\bar{\Lambda}$s created in the corona $\mathcal{P}^{\Lambda/\bar{\Lambda}}_{REC}$, that were taken in the original approach as unpolarized. 

In our previous works~\cite{Ayala:2020soy,Ayala:2021xrn} we have expressed the polarizations as functions of the number of $\Lambda$s produced in the core and in the corona, $N_{\Lambda\ {\mbox{\tiny{QGP}}}}$, $N_{\Lambda\ {\mbox{\tiny{REC}}}}$, respectively, as
\begin{eqnarray}
\mathcal{P}^\Lambda&=&\frac{ \mathcal{P}^\Lambda_{REC} + z\frac{
N_{\Lambda\ {\mbox{\tiny{QGP}}}} }{N_{\Lambda\ {\mbox{\tiny{REC}}}}}}{ \left( 1 + \frac{N_{\Lambda\ {\mbox{\tiny{QGP}}}}}{N_{\Lambda\ {\mbox{\tiny{REC}}}}}\right)},\nonumber\\
\mathcal{P}^{\overline{\Lambda}}&=&\frac{ \mathcal{P}^{\overline{\Lambda}}_{REC} + \bar{z}\left(\frac{w'}{w}\right)
\frac{N_{\Lambda\ {\mbox{\tiny{QGP}}}} }{N_{\Lambda\ {\mbox{\tiny{REC}}}}}}{ \left( 1 + 
\left(\frac{w'}{w}\right)
\frac{N_{\Lambda\ {\mbox{\tiny{QGP}}}}}{N_{\Lambda\ {\mbox{\tiny{REC}}}}}\right)},
\label{eq:pola}
\end{eqnarray}
where the ratio of the number of $\bar{\Lambda}$s and $\Lambda$s, in the core and in the corona are 
$w' = N_{\overline{\Lambda}\ {\mbox{\tiny{QGP}}}}/N_{\Lambda\ {\mbox{\tiny{QGP}}}}$ and 
$w' = N_{\overline{\Lambda}\ {\mbox{\tiny{REC}}}}/N_{\Lambda\ {\mbox{\tiny{REC}}}}$,  respectively,
and $z$ and $\overline{z}$ are the intrinsic polarizations for $\Lambda$ and $\overline{\Lambda}$, respectively~\cite{Ayala:2020ndx,Ayala:2019iin}. In this work we propose a way to estimate the contribution to the global polarization from $\Lambda$s and $\bar{\Lambda}$s created in the corona, as well as the feasibility to measure these quantities using the MPD experiment~\cite{MPD:2022qhn}.  

The angular distribution of protons produced in the weak decays of $\Lambda$'s, which are used to experimentally measure the polarization, is given by 
\begin{equation}
    \label{eq:angdist}
    \frac{dN}{d\Omega} = \frac{N}{4\pi}\left( 1 + \alpha \mathcal{P} \cos{\Theta^{*}}\right)
\end{equation}
where $\alpha \approx 0.732$\cite{BESIII:2018cnd} is the decay asymmetry parameter, $\mathcal{P}$ is the $\Lambda$ polarization and $\Theta^{*}$ is the angle between the proton and total angular momentum direction, measured in the $\Lambda$ rest frame. The work is organized as follows: In Sec.~\ref{sec2}, we use the set-up provided by Eq.~(\ref{eq:angdist}) to estimate the contribution to global polarization from $\Lambda$s produced in the corona. In Sec.~\ref{sec3}, we study the implementation of this estimate within the MPD framework to determine the experimental feasibility measurement. We finally summarize and discuss our results in Sec.~\ref{concl}.

\section{Hyperons from the corona and their contribution to the global polarization}\label{sec2}

The core-corona model assumes that in peripheral heavy-ion collisions, when the critical density of participants to produce a QGP is barely or not achieved, particles are instead produced by nucleon-nucleon interactions. Consequently, the polarization of $\Lambda$ hyperons is produced during the hadronization process by an as yet unknown mechanism, for instance, the DeGrand--Miettinen spin precession mechanism~\cite{DeGrand:1980gc,DeGrand:1981pe,Fujita:1988fr,Kitsukawa:2000zk}), Although possibly small, this polarization may not completely be ignored. Then, its projection along the direction of the total angular momentum could be different from zero, therefore contributing to the global polarization.

It is well-know that the $\Lambda$'s transverse polarization $\mathcal{P}_{T}$ in p + p collisions is different from zero and that in average it can take on values between -10\% and -40\%. This is summarized in table~\ref{tab:my_label} for different collision energies.
\begin{table}[H]
    \centering
    \begin{tabular}{c|c}
        $\sqrt{s}$ (GeV)& $\mathcal{P}$ \\
         19.6 & -0.25 $\pm$ 0.26 \\
         26.0 & -0.24 $\pm$ 0.09 \\
         53   & -0.34 $\pm$ 0.07 \\
         62   & -0.40 $\pm$ 0.20
    \end{tabular}
\caption{Average transverse polarization measured in p + p collisions~\cite{R608:1986ltk,Blobel:1977ms,EHS-RCBC:1984bxo,Jaeger:1974in}.}
    \label{tab:my_label}
\end{table}

Assuming that the $\Lambda$s produced in the corona show a similar transverse polarization with respect to its production plane, we project this polarization along the global polarization direction and estimate whether its contribution is different from zero.

First, we consider that each $\Lambda$ is produced in a p + p collision from the participants in the corona and that its polarization points along the direction of the production plane, which is defined by the direction of the incoming proton $\vec{p}_{beam}$ and the $\Lambda$ direction, $\vec{p}_{\Lambda}$ 

    \begin{equation}
        \hat{n} \equiv \frac{\vec{p}_{beam}\times \vec{p}_{\Lambda}}{|\vec{p}_{beam}\times \vec{p}_{\Lambda}|}
    \end{equation}
Assuming that the beam direction is parallel to $\hat{z}$, we can express $\hat{n}$ in terms of the transverse momentum components of the produced $\Lambda$, namely,
    \begin{equation}
        \hat{n} = \frac{1}{p_{T_{\Lambda}}}(-p_{y_{\Lambda}},p_{x_{\Lambda}},0).
    \label{eq:hatn}
    \end{equation}
For the sake of simplicity we consider that the polarization $\mathcal{P}_{REC}$ is only different from zero along $\hat{n}$. Therefore, its 
contribution to the global polarization can be determined using Eq.~(\ref{eq:angdist}) written as
    \begin{equation}
        \frac{dN}{d\Omega} = \frac{N}{4\pi}\left( 1 + \alpha \mathcal{P}_{REC} \cos{\sigma^{*}}\right),
    \end{equation}
    where $\sigma^{*}$ is the angle between $\hat{n}$ and the direction of the angular momentum $\hat{L} = \hat{b}\times\hat{p}_{beam} = (\sin{\Psi_{RP}},-\cos{\Psi_{RP}},0)$.
    
    Therefore $\cos \sigma^{*}$ is given by
    \begin{equation}
        \cos{\sigma^{*}} = \hat{n} \cdot \hat{L} = \frac{1}{p_{T_{\Lambda}}}\left( -p_{y_{\Lambda}} \sin{\Psi_{RP}} - p_{x_{\Lambda}} \cos{\Psi_{RP}}\right).
    \end{equation}
    Substituting 
    \begin{eqnarray}
    p_{x_{\Lambda}} &=& p_{\Lambda}\sin{\theta_{\Lambda}\cos{\phi_{\Lambda}}} \nonumber \\
    p_{y_{\Lambda}} &=& p_{\Lambda}\sin{\theta_{\Lambda}\sin{\phi_{\Lambda}}} \nonumber \\
    p_{T_{\Lambda}} &=& p_{\Lambda}\sin{\theta_{\Lambda}}
    \end{eqnarray}
we obtain
    \begin{eqnarray}
        \cos{\sigma^{*}} &=& -\sin{\phi_{\Lambda}}\sin{\Psi_{RP}} - \cos{\phi_{\Lambda}}\cos{\Psi_{RP}} \nonumber \\
        &=& - \cos{(\phi_{\Lambda}-\Psi_{RP}).}
    \end{eqnarray}
The angular distribution can be written as
    \begin{equation}
        \frac{dN}{d\Omega} = \frac{N}{4\pi}\left( 1 - \alpha\mathcal{P}_{T} \cos{(\phi_{\Lambda}-\Psi_{RP})} \right).
    \end{equation}
Integrating over the polar angle $\theta$, we get
\begin{eqnarray}
    \frac{dN}{d\phi}&=&\int_0^{\pi}\left[ \frac{N}{4\pi}(1-\alpha\mathcal{P}_T \cos{(\phi_{\Lambda}-\Psi_{RP})} )\right] \sin{\theta}d\theta \nonumber \\
    &=& \frac{N}{2\pi}(1 - \alpha \mathcal{P}_T \cos{(\phi_{\Lambda} - \Psi_{RP})}),
\end{eqnarray}
from where we can compute the mean angular distribution $\langle \cos{(\phi_{\Lambda}-\Psi_{RP})} \rangle$, which is given by
\begin{equation}
    \langle \cos{(\phi_{\Lambda}-\Psi_{RP})} \rangle = - \frac{\alpha \mathcal{P}_T}{2}.
\end{equation}
The transverse polarization projected along the angular momentum is thus given by
\begin{equation}
    \mathcal{P}_T = \frac{- 2  \langle \cos{(\phi_{\Lambda}-\Psi_{RP})} \rangle}{\alpha},
    \label{eqpol}
\end{equation}
which differs from the global polarization given by 
\begin{equation}
\mathcal{P}_{\Lambda} = - \frac{8  \langle \sin{(\phi_{p}-\Psi_{RP})} \rangle}{\pi \alpha}.
\end{equation}
Notice that the right-hand side of Eq.~\eqref{eqpol}, expressing the average transverse $\Lambda$ polarization from processes in the corona, appears also in the expression to determine the directed flow. Therefore, a non-vanishing result for this average cannot be exclusively attributed  to a non-vanishing polarization. Nevertheless, it has been experimentally determined that the directed flow has very distinct characteristics~\cite{STAR:2018gji,Gardim:2011qn ,ALICE:2013xri,Luzum:2010fb, Teaney:2010vd}. For example, although its odd component tends to zero when the pseudorapidity $\eta$ goes to zero, changing sign as a function of both $\eta$ and the transverse momentum $p_t$, the existence of a non-vanishing even component 
produces that the above mentioned average is also non-vanishing. The latter is essentially $\eta$-independent and originates in the fluctuations of the initial collision geometry and on the subsequent hydrodynamical evolution which produces that this component of the directed flow achieves values up to 10\%. Therefore, if the directed flow is non-vanishing, one can expect, in agreement with Eq.~(\ref{eqpol}), that the contribution to global $\Lambda$ polarization in the corona is also non-vanishing.

\section{Implementation within the MPD framework}\label{sec3}

We generate a small MC sample consisting of 150,000 Bi + Bi events at $\sqrt{s_{NN}}=9.2$ GeV using the UrQMD event generator~\cite{Bass:1998ca,Bleicher:1999xi}, which allows a core-corona separation~\cite{Steinheimer:2011mp}. The generator implements a hybrid model for the hot and dense stage~\cite{Petersen:2008dd} that includes a fluid-dynamical evolution carried out by SHASTA (Sharp and Smooth Transport Algorithm)~\cite{Rischke:1995ir,Rischke:1995mt}. For the simulation we choose an equation of state that includes deconfinement plus a chiral phase transition. The UrQMD generator considers that particles from the core are those that come from regions above a given quark density in a given $\eta$ window.

We use the 
MpdRoot framework to reconstruct $\Lambda$ decay products and assign them a polarization, in the same fashion as is done for the dedicated analysis to measure global polarization in the MPD~\cite{keyNaz}. Due to the size of the sample, we assign a transverse polarization of 40\% for the $\Lambda$s in the corona. The direction of this polarization is parallel to the vector $\hat n$ defined in Eq.~(\ref{eq:hatn}). To avoid distortion of the results by efficiency reconstruction within detector, at this level of the analysis we use only the MC information.

\begin{figure}[H]
    \centering
    \includegraphics[width=\linewidth]{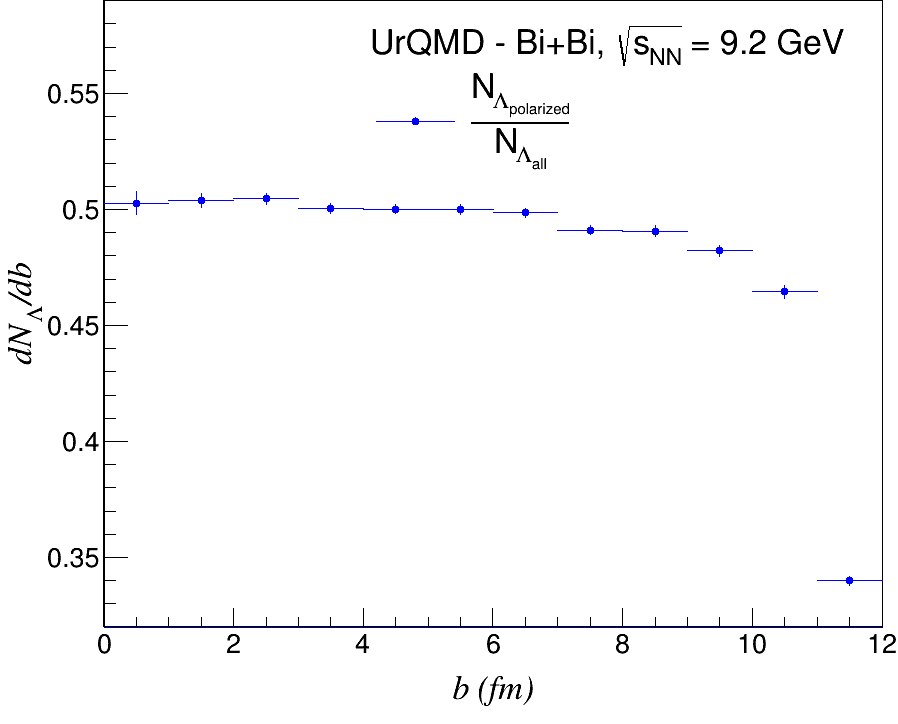}
    \caption{Ratio between polarized $\Lambda$s produced in the corona with respect all $\Lambda$s, including those secondary produced by decays.}
    \label{fig:ratio}
\end{figure}

Although the generator produces a different proportion between the number of particles in the core and in the corona with respect to the Glauber model used in our previous calculations, this analysis sheds light on the possibility to measure the conversion of local transverse into global $\Lambda$ polarization. Using this procedure, we obtain a final sample consisting of approximately half of the generated $\Lambda$s showing some polarization, as is shown in Fig.~\ref{fig:ratio}. 

\begin{figure}[H]
    \includegraphics[width=\linewidth]{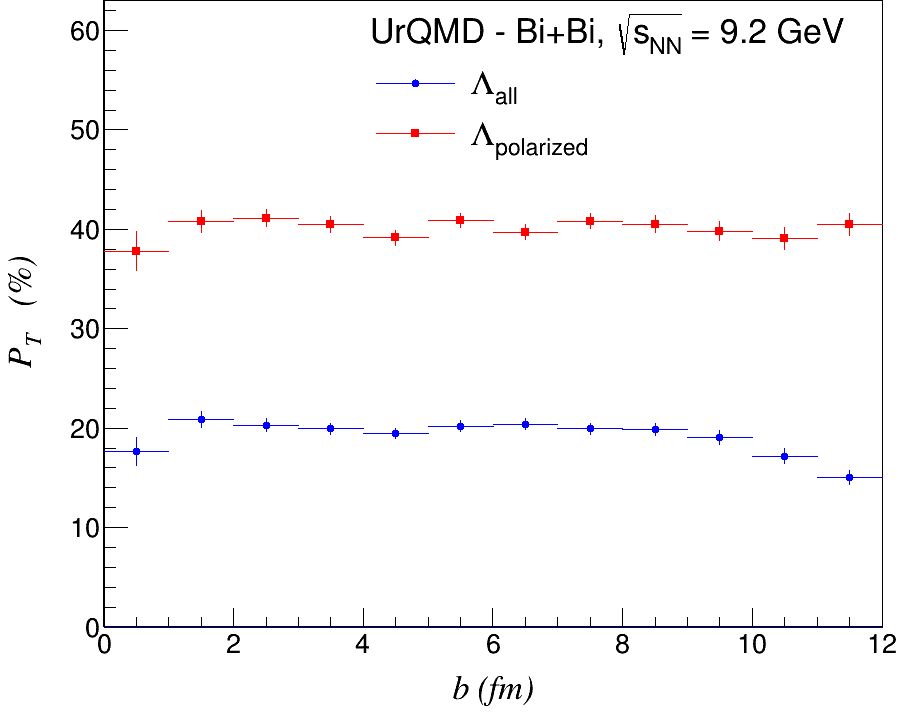}
    \caption{Transverse polarization of $\Lambda$s from the corona. The polarization is 40\% and it is diluted by the contribution from all of the produced $\Lambda$s.}
    \label{fig:transpol}
\end{figure}

To test the procedure for the assignment of the polarization, we determine the transverse polarization using the angular distribution given by Eq.~(\ref{eq:angdist}). The transverse polarization for the $\Lambda$s produced in the corona is shown in Fig.~\ref{fig:transpol} together with the transverse polarization of the whole sample. We notice that the transverse polarization is diluted by the unpolarized $\Lambda$s, and that this dilution corresponds to the proportion of unpolarized particles. 

\begin{figure}[H]
    \includegraphics[width=\linewidth]{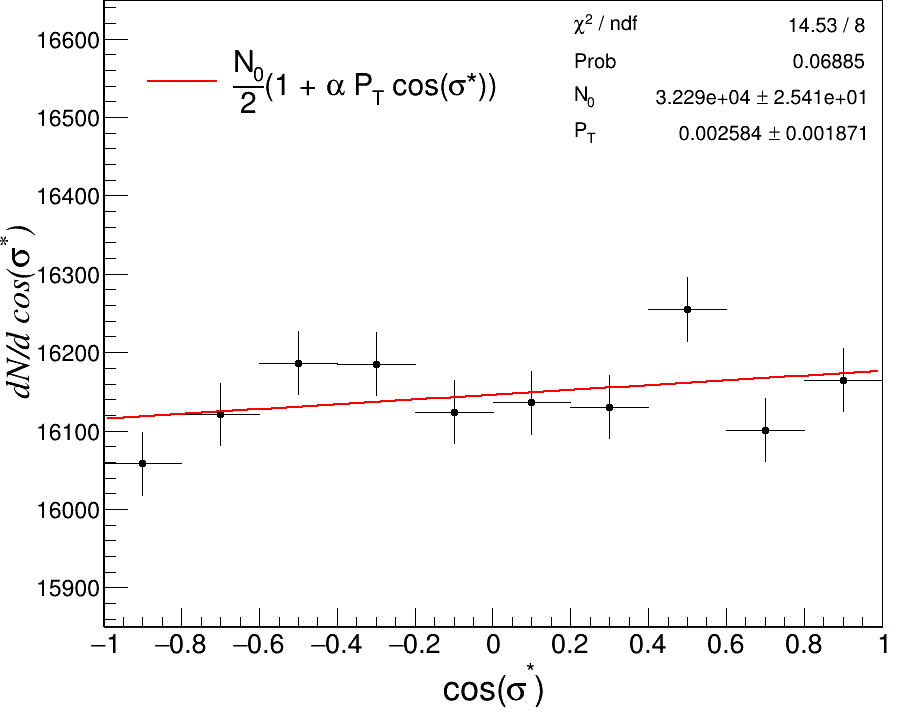}\\
    \includegraphics[width=\linewidth]{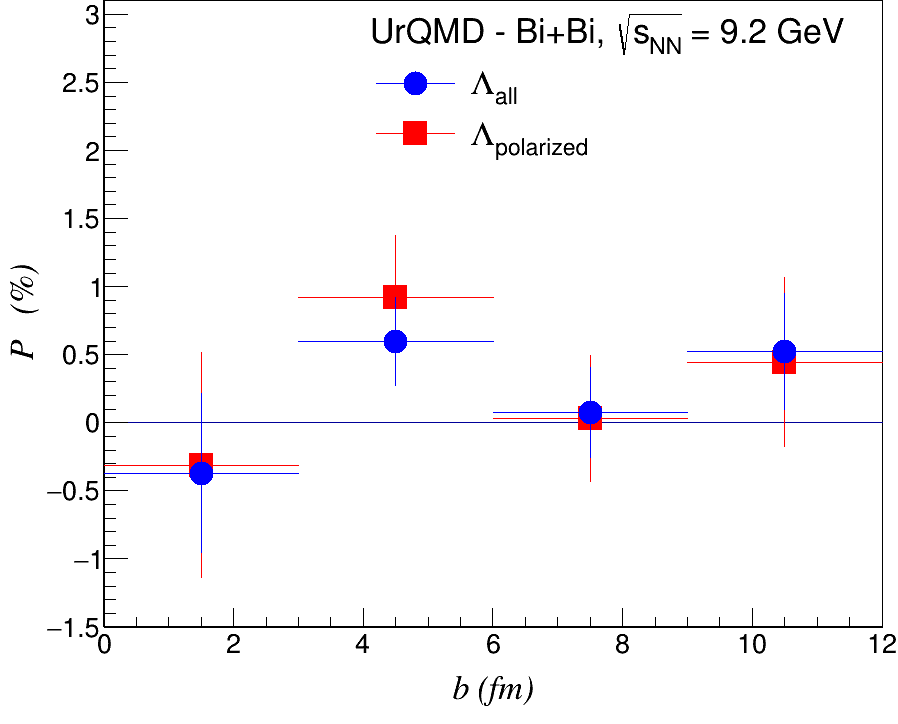}
    \caption{Global  $\Lambda$ polarization from the corona. The upper plot shows the angular distribution for all the sample. The lower plot shows the polarization as a function of the impact parameter.  The polarization is of order of 1\%, and it is diluted by the contribution from all the produced $\Lambda$s. More statistics is needed to have conclusive results.}
    \label{fig:globpol}
\end{figure}

Following a similar procedure, we calculate the contribution to the global polarization. The results are  shown in the Fig.~\ref{fig:globpol}. In average the polarization is small ($\sim $0.25\%) but different from zero. However, as a function of the impact parameter $b$, it looks to be larger for some bins. Nevertheless, the results are still not conclusive due the lack of statistics. Improvements on this analysis constitute work is in progress which will be reported elsewhere.

\section{Summary}\label{concl}

We have implemented a description of the global $\Lambda$ polarization within the core-corona model, which gives a good account of experimental data for the HADES, STAR and NICA energy range. The
UrQMD event generator has been used to simulate both the hydrodynamical phase of the core and the cascade transport in the corona, as well as the separate contributions from these regions to the $\Lambda$ polarization.
MpdRoot has been used to simulate the decay and transport of the polarization of the $\Lambda$ decay products.
We have shown that a local polarization could contribute to the global polarization. This has been implemented by allowing that $\Lambda$s from the corona are produced with a fraction of the transverse polarization that is measured in p + p collisions. The existence of a non-vanishing even component of the directed flow can also be linked to a non-vanishing average transverse polarization. However, the results are still inconclusive due to the small size of the sample. A full analysis is on its way. This is work in progress and the results will soon be reported elsewhere.

\section*{Acknowledgments}

The simulations were produced within the ICN-UNAM cluster and on the basis of the HybriLIT heterogeneous computing platform (LIT, JINR) \hyperref[http://hlit.jinr.ru]{http://hlit.jinr.ru}. Support for this work was received in part by UNAM-PAPIIT IG100322 and by Consejo Nacional de Ciencia y Tecnolog\'ia grant number A1-S-7655. METY is grateful for the hospitality of Perimeter Institute where part of this work was carried out.

\end{multicols}

\medline
\begin{multicols}{2}
%
\nocite{*}
\bibliographystyle{rmf-style}
\bibliography{ref}

\begin{thebibliography}{99}
\providecommand{\url}[1]{\texttt{#1}}
\providecommand{\urlprefix}{URL }
\expandafter\ifx\csname urlstyle\endcsname\relax
  \providecommand{\doi}[1]{\discretionary{}{}{}#1}\else
  \providecommand{\doi}{\discretionary{}{}{}\begingroup \urlstyle{rm}\Url}\fi

\bibitem{Ayala:2001jp}
A.~Ayala, et~al.,
\newblock {$\Lambda^{0}$ polarization as a probe for production of deconfined
  matter in ultrarelativistic heavy ion collisions},
\newblock Phys. Rev. C 65 (2002) 024902,
\newblock \url{https://doi.org/10.1103/PhysRevC.65.024902}

\bibitem{STAR:2017ckg}
L.~Adamczyk et~al.,
\newblock {Global $\Lambda$ hyperon polarization in nuclear collisions:
  evidence for the most vortical fluid},
\newblock Nature 548 (2017) 62,
\newblock \url{https://doi.org/10.1038/nature23004}

\bibitem{STAR:2021beb}
M.~S. Abdallah et~al.,
\newblock {Global $\Lambda$-hyperon polarization in Au+Au collisions at $\sqrt
  {s_{NN}}$=3~GeV},
\newblock Phys. Rev. C 104 (2021) L061901,
\newblock \url{https://doi.org/10.1103/PhysRevC.104.L061901}

\bibitem{HADES:2022enx}
R.~Abou~Yassine et~al.,
\newblock {Measurement of global polarization of \ensuremath{\Lambda} hyperons
  in few-GeV heavy-ion collisions},
\newblock Phys. Lett. B 835 (2022) 137506,
\newblock \url{https://doi.org/10.1016/j.physletb.2022.137506}

\bibitem{Ayala:2020soy}
A.~Ayala et~al.,
\newblock {Core meets corona: A two-component source to explain $\Lambda$ and
  $\bar \Lambda$ global polarization in semi-central heavy-ion collisions},
\newblock Phys. Lett. B 810 (2020) 135818,
\newblock \url{https://doi.org/10.1016/j.physletb.2020.135818}

\bibitem{Ayala:2021xrn}
A.~Ayala, et~al.,
\newblock {Rise and fall of $\Lambda$ and $\bar{\Lambda}$ global polarization
  in semi-central heavy-ion collisions at HADES, NICA and RHIC energies from
  the core-corona model},
\newblock Phys. Rev. C 105 (2022) 034907,
\newblock \url{https://doi.org/10.1103/PhysRevC.105.034907}

\bibitem{Becattini:2008ya}
F.~Becattini and J.~Manninen,
\newblock {Centrality dependence of strangeness production in heavy-ion
  collisions as a geometrical effect of core-corona superposition},
\newblock Phys. Lett. B 673 (2009) 19,
\newblock \url{https://doi.org/10.1016/j.physletb.2009.01.066}

\bibitem{Aichelin:2010ed}
J.~Aichelin and K.~Werner,
\newblock {Is the centrality dependence of the elliptic flow $v_2$ and of the
  average $ < p_T > $ more than a Core-Corona Effect?},
\newblock Phys. Rev. C 82 (2010) 034906,
\newblock \url{https://doi.org/10.1103/PhysRevC.82.034906}

\bibitem{Ayala:2022yyx}
A.~Ayala, et~al.,
\newblock {$\Lambda$ and $\bar{\Lambda}$ global polarization from the
  core-corona model},
\newblock Rev. Mex. Fis. Suppl. 3 (2022) 040914,
\newblock \url{https://doi.org/10.31349/SuplRevMexFis.3.040914}

\bibitem{Ayala:2020ndx}
A.~Ayala, et~al.,
\newblock {Relaxation time for the alignment between the spin of a finite-mass
  quark or antiquark and the thermal vorticity in relativistic heavy-ion
  collisions},
\newblock Phys. Rev. D 102 (2020) 056019,
\newblock \url{https://doi.org/10.1103/PhysRevD.102.056019}

\bibitem{Ayala:2019iin}
A.~Ayala, et~al.,
\newblock {Relaxation time for quark spin and thermal vorticity alignment in
  heavy-ion collisions},
\newblock Phys. Lett. B 801 (2020) 135169,
\newblock \url{https://doi.org/10.1016/j.physletb.2019.135169}

\bibitem{MPD:2022qhn}
V.~Abgaryan et~al.,
\newblock {Status and initial physics performance studies of the MPD experiment
  at NICA},
\newblock Eur. Phys. J. A 58 (2022) 140,
\newblock \url{https://doi.org/10.1140/epja/s10050-022-00750-6}

\bibitem{BESIII:2018cnd}
M.~Ablikim et~al.,
\newblock {Polarization and Entanglement in Baryon-Antibaryon Pair Production
  in Electron-Positron Annihilation},
\newblock Nature Phys. 15 (2019) 631,
\newblock \url{https://doi.org/10.1038/s41567-019-0494-8}

\bibitem{DeGrand:1980gc}
T.~A. DeGrand and H.~I. Miettinen,
\newblock {Quark Dynamics of Polarization in Inclusive Hadron Production},
\newblock Phys. Rev. D 23 (1981) 1227,
\newblock \url{https://doi.org/10.1103/PhysRevD.23.1227}

\bibitem{DeGrand:1981pe}
T.~A. DeGrand and H.~I. Miettinen,
\newblock {Models for Polarization Asymmetry in Inclusive Hadron Production},
\newblock Phys. Rev. D 24 (1981) 2419,
\newblock \url{https://doi.org/10.1103/PhysRevD.24.2419}

\bibitem{Fujita:1988fr}
T.~Fujita and T.~Matsuyama,
\newblock {A Comment on the Degrand-miettinen Model for the Polarization of
  $\Lambda$ in Proton Proton Collisions},
\newblock Phys. Rev. D 38 (1988) 401,
\newblock \url{https://doi.org/10.1103/PhysRevD.38.401}

\bibitem{Kitsukawa:2000zk}
Y.~Kitsukawa and K.~Kubo,
\newblock {Hadron spin polarization produced by a dynamical spin-orbit
  interaction},
\newblock Prog. Theor. Phys. 103 (2000) 1173,
\newblock \url{https://doi.org/10.1143/PTP.103.1173}

\bibitem{R608:1986ltk}
A.~M. Smith et~al.,
\newblock {$\Lambda^0$ Polarization in Proton Proton Interactions From
  $\sqrt{s} = 31$-{GeV} to 62-{GeV}},
\newblock Phys. Lett. B 185 (1987) 209,
\newblock \url{https://doi.org/10.1016/0370-2693(87)91556-5}

\bibitem{Blobel:1977ms}
V.~Blobel, et~al.,
\newblock {Transverse Momentum Dependence in Proton Proton Interactions at
  24-GeV/c},
\newblock Nucl. Phys. B 122 (1977) 429,
\newblock \url{https://doi.org/10.1016/0550-3213(77)90137-7}

\bibitem{EHS-RCBC:1984bxo}
M.~Asai et~al.,
\newblock {Inclusive $K^0_S$, $\Lambda$ and $\bar{\Lambda}$ Production in
  360-{GeV}/$c p p$ Interactions Using the European Hybrid Spectrometer},
\newblock Z. Phys. C 27 (1985) 11,
\newblock \url{https://doi.org/10.1007/BF01642475}

\bibitem{Jaeger:1974in}
K.~Jaeger, et~al.,
\newblock {Characteristics of V0 and gamma Production in p p Interactions at
  205-GeV/c},
\newblock Phys. Rev. D 11 (1975) 2405,
\newblock \url{https://doi.org/10.1103/PhysRevD.11.2405}

\bibitem{STAR:2018gji}
J.~Adam et~al.,
\newblock {Beam energy dependence of rapidity-even dipolar flow in Au+Au
  collisions},
\newblock Phys. Lett. B 784 (2018) 26,
\newblock \url{https://doi.org/10.1016/j.physletb.2018.07.013}

\bibitem{Gardim:2011qn}
F.~G. Gardim, et~al.,
\newblock {Directed flow at mid-rapidity in event-by-event hydrodynamics},
\newblock Phys. Rev. C 83 (2011) 064901,
\newblock \url{https://doi.org/10.1103/PhysRevC.83.064901}

\bibitem{ALICE:2013xri}
B.~Abelev et~al.,
\newblock {Directed Flow of Charged Particles at Midrapidity Relative to the
  Spectator Plane in Pb-Pb Collisions at $\sqrt{s_{NN}}$=2.76 TeV},
\newblock Phys. Rev. Lett. 111 (2013) 232302,
\newblock \url{https://doi.org/10.1103/PhysRevLett.111.232302}

\bibitem{Luzum:2010fb}
M.~Luzum and J.-Y. Ollitrault,
\newblock {Directed flow at midrapidity in heavy-ion collisions},
\newblock Phys. Rev. Lett. 106 (2011) 102301,
\newblock \url{https://doi.org/10.1103/PhysRevLett.106.102301}

\bibitem{Teaney:2010vd}
D.~Teaney and L.~Yan,
\newblock {Triangularity and Dipole Asymmetry in Heavy Ion Collisions},
\newblock Phys. Rev. C 83 (2011) 064904,
\newblock \url{https://doi.org/10.1103/PhysRevC.83.064904}

\bibitem{Bass:1998ca}
S.~A. Bass et~al.,
\newblock {Microscopic models for ultrarelativistic heavy ion collisions},
\newblock Prog. Part. Nucl. Phys. 41 (1998) 255,
\newblock \url{https://doi.org/10.1016/S0146-6410(98)00058-1}

\bibitem{Bleicher:1999xi}
M.~Bleicher et~al.,
\newblock {Relativistic hadron hadron collisions in the ultrarelativistic
  quantum molecular dynamics model},
\newblock J. Phys. G 25 (1999) 1859,
\newblock \url{10.1088/0954-3899/25/9/308}

\bibitem{Steinheimer:2011mp}
J.~Steinheimer and M.~Bleicher,
\newblock {Core-corona separation in the UrQMD hybrid model},
\newblock Phys. Rev. C 84 (2011) 024905,
\newblock \url{https://doi.org/10.1103/PhysRevC.84.024905}

\bibitem{Petersen:2008dd}
H.~Petersen, et~al.,
\newblock {A Fully Integrated Transport Approach to Heavy Ion Reactions with an
  Intermediate Hydrodynamic Stage},
\newblock Phys. Rev. C 78 (2008) 044901,
\newblock \url{https://doi.org/10.1103/PhysRevC.78.044901}

\bibitem{Rischke:1995ir}
D.~H. Rischke, S.~Bernard, and J.~A. Maruhn,
\newblock {Relativistic hydrodynamics for heavy ion collisions. 1. General
  aspects and expansion into vacuum},
\newblock Nucl. Phys. A 595 (1995) 346,
\newblock \url{https://doi.org/10.1016/0375-9474(95)00355-1}

\bibitem{Rischke:1995mt}
D.~H. Rischke, Y.~Pursun, and J.~A. Maruhn,
\newblock {Relativistic hydrodynamics for heavy ion collisions. 2. Compression
  of nuclear matter and the phase transition to the quark - gluon plasma},
\newblock Nucl. Phys. A 595 (1995) 383,
\newblock \url{https://doi.org/10.1016/0375-9474(95)00356-3}

\bibitem{keyNaz}
E.~Nazarova et~al.,
\newblock Study of hyperon global polarization at MPD (2022),
\newblock
  \urlprefix\url{http://indico.oris.mephi.ru/event/298/session/1/contribution/12},
\newblock Workshop on physics performance studies at NICA (NICA-2022).

\bibitem{2022137003}
M.~Abdallah, et~al.,
\newblock Disappearance of partonic collectivity in $\sqrt{s_{NN}} = 3$GeV
  Au+Au collisions at RHIC,
\newblock Physics Letters B 827 (2022) 137003,
\newblock \url{https://doi.org/10.1016/j.physletb.2022.137003}

\end{thebibliography}

\end{multicols}
\end{document}